\documentclass{article}
\usepackage{amsmath,amsfonts,amssymb}
\usepackage{calc,graphicx,color}
\usepackage[width=372pt,height=588pt,top=126pt]{geometry} % IOP journal standard textwidth and text height
\usepackage[hyperindex,breaklinks]{hyperref}
\usepackage[numeric]{amsrefs} % Tested with version 1.23, 2002/02/28.

\newcommand{\figtobox}[3][{}]{%
	\expandafter\newsavebox\csname #2box\endcsname%
	\expandafter\savebox\csname #2box\endcsname{#1\input{#3.pstex_t}}%
	\expandafter\def\csname #2\endcsname%
		{\setlength{\figoff}{1.0ex-.5\expandafter\ht\csname #2box\endcsname}%
			\raisebox{\figoff}{\expandafter\usebox\csname #2box\endcsname}}%
}
\newlength{\figoff}

\figtobox[\def\a{$a$}\def\b{$b$}\def\c{$c$}\def\d{$d$}\def\e{$e$}\def\f{$f$}]{tet}{tet}
\figtobox[\def\a{$a$}\def\b{$b$}\def\c{$c$}\def\d{$d$}\def\e{$e$}]{YYlow}{4-vert-split}
\figtobox[\def\a{$a$}\def\b{$b$}\def\c{$c$}\def\d{$d$}\def\e{$e$}]{YY}{4-vert-split}
\figtobox[\def\a{$a$}\def\b{$b$}\def\c{$c$}\def\d{$d$}\def\f{$f$}]{YYh}{4-vert-splith}
\figtobox[\def\a{$a$}\def\b{$b$}\def\c{$c$}]{thabc}{theta}
\figtobox[\def\a{$a$}\def\b{$d$}\def\c{$e$}]{thade}{theta}
\figtobox[\def\a{$b$}\def\b{$c$}\def\c{$e$}]{thbce}{theta}
\figtobox[\def\j{$j$}]{bubj}{bubble}
\figtobox[\def\j{$j$}]{vbubf}{vbubble}
\figtobox[\def\j{$j_1$}]{vbubei}{vbubble}
\figtobox[\def\j{$j_2$}]{vbubeii}{vbubble}
\figtobox[\def\j{$j_3$}]{vbubeiii}{vbubble}
\figtobox[\def\j{$j_4$}]{vbubeiv}{vbubble}
\figtobox[\def\a{$j_1$}\def\b{$j_2$}\def\c{$j_3$}\def\d{$j_4$}]{eye}{eye}
\figtobox[\def\a{$a$}\def\b{$b$}\def\c{$c$}\def\d{$d$}]{bcv}{4-vert}
\figtobox[%
		\def\a{$0$}%
		\def\b{$1$}%
		\def\c{$2$}%
		\def\d{$3$}%
		\def\e{$4$}%
		\def\joa{$j_{1,0}$}%
		\def\job{$j_{1,1}$}%
		\def\joc{$j_{1,2}$}%
		\def\jod{$j_{1,3}$}%
		\def\joe{$j_{1,4}$}%
		\def\jia{$j_{2,0}$}%
		\def\jib{$j_{2,1}$}%
		\def\jic{$j_{2,2}$}%
		\def\jid{$j_{2,3}$}%
		\def\jie{$j_{2,4}$}%
	]{tenjv}{10j-cros}
\figtobox[%
		\def\a{}%
		\def\b{}%
		\def\c{(a)}%
		\def\d{}%
		\def\e{}%
		\def\joa{}%
		\def\job{}%
		\def\joc{}%
		\def\jod{}%
		\def\joe{}%
		\def\jia{}%
		\def\jib{}%
		\def\jic{}%
		\def\jid{}%
		\def\jie{}%
	]{tenj}{10j-cros}
\figtobox[\def\a{(b)}]{fifj}{15j-cros}
\figtobox[\def\a{(c)}]{ladder}{ladder}
\figtobox[\def\a{(d)}]{bubbles}{bubbles}
\figtobox[\def\j{$j$}]{curl}{curl}
\figtobox[\def\j{$j$}]{line}{line}
\figtobox[\def\a{$a$}\def\b{$b$}\def\c{$c$}]{tvert}{3-vert}
\figtobox[\def\a{$a$}\def\b{$b$}\def\c{$c$}]{tvertb}{3-vert-braid}
\figtobox[\def\a{$a$}\def\b{$b$}\def\c{$c$}]{tvertt}{3-vert-dot}
\figtobox[\def\a{$a$}\def\b{$b$}\def\c{$c$}]{tvertbt}{3-vert-braid-dot}
\newcommand{\TetOp}{\operatorname{Tet}}
\newcommand{\Tet}[1]{\ensuremath{\TetOp\left[\begin{matrix}#1\end{matrix}\right]}}

\newcommand{\ev}[1]{\left\langle{#1}\right\rangle}
\newcommand{\qhi}[1]{\left\lfloor{#1}\right\rceil} % quantum half-integer
\def\nto{\nrightarrow}
\def\tr{\operatorname{tr}}
\def\lb{\operatorname{lb}}
\def\ub{\operatorname{ub}}
\def\oo{\infty}
\def\C{\mathbb{C}}
\def\R{\mathbb{R}}

\def\eps{\varepsilon}
\def\suu{\mathfrak{su}(2)}
\def\suq{\mathfrak{su}_q(2)}
\def\suqp{\mathfrak{su}_{q'}(2)}
\def\suqi{\mathfrak{su}_{q^{-1}}(2)}
\def\isom{\cong}
\def\Spin{\mathrm{Spin}}
\def\spin{\mathfrak{spin}}
\def\dist{\mathrm{dist}}

\title{$q$-deformed spin foam models of quantum gravity}
\author{Igor Khavkine$^1$ and J. Daniel Christensen$^2$\\
	\small
	$^1$ Department of Applied Mathematics,
		University of Western Ontario, London, Ontario, Canada\\
	\small
	$^2$ Department of Mathematics,
		University of Western Ontario, London, Ontario, Canada\\
	\small
	E-mail: ikhavkin@uwo.ca and jdc@uwo.ca}

\begin{document}
\maketitle
\begin{abstract}
	We numerically study Barrett-Crane models of Riemannian quantum
	gravity. We have extended the existing numerical techniques to handle
	$q$-deformed models and arbitrary space-time triangulations.
	We present and interpret expectation values of a few selected
	observables for each model, including a spin-spin correlation
	function which gives insight into the behaviour of the models.
	We find the surprising result that, as the deformation parameter $q$
	goes to $1$ through roots of unity, the limit is discontinuous. 
	\vspace{1ex}

\end{abstract}

\section{Introduction}
Spin foam models were first introduced as a space-time alternative to
the spin network description of states in loop quantum
gravity~\cite{Baez-spinfoam}. The most studied spin foam models are due
to Barrett and Crane~\cites{BC-riem,BC-lor}. A spin foam is a
discretization of space-time where the fundamental degrees of freedom
are the areas labelling its $2$-dimensional faces.

An important goal in the investigation of spin foam models is to obtain
predictions that can be compared to the large scale, classical, or
semiclassical behavior of gravity. This work
continues the numerical investigation of the
physical properties of spin foam models of Riemannian quantum gravity begun
in~\cites{CE,BC-pos,BCHT,BCE}.
In this paper, we extend the computations to the $q$-deformed
Barrett-Crane model and to larger space-time triangulations.

The main applications of $q$-deformation are two-fold. On the one hand,
it can act as a regulator for divergent models, as is apparent in the
link between the Ponzano-Regge~\cite{PR} and Turaev-Viro~\cite{TV}
models. On the other hand, Smolin~\cite{Smolin-posCC} has argued that
$q$-deformation is necessary to account for a positive cosmological
constant. Both of these aspects are explored in more detail in
Section~\ref{qdef-appl}. A surprising result of our work is evidence
that the limit, as the cosmological constant is taken to zero through
positive values, is discontinuous.

Large triangulations are necessary to approximate semiclassical
space-times. The possibility of obtaining numerical results from larger
triangulations takes us one step closer to that goal and increases the
number of facets from which the physical properties of a spin foam model
may be examined.  As an example, we are able to study how the
spin-spin correlation varies with the distance between faces in the
triangulation. 

This paper is structured as follows. We begin in Section~\ref{qdef} by reviewing
the basics of $q$-deformation and discussing in detail its aforementioned
applications. Section~\ref{bc-deform} reviews the details of the
Barrett-Crane model, summarizes the necessary changes for its
$q$-deformation, and defines several observables associated to spin
foams. In Section~\ref{numsim}, we review the existing numerical simulation
techniques and how they need to be generalized to
handle $q$-deformation and larger triangulations.
Section~\ref{results} presents the results of our numerical simulations.
In Section~\ref{se:conclusion}, we give our conclusions and list some
avenues for future research.
The Appendix briefly summarizes our notational conventions and useful
formulas.

\section{Deformation of $\suu$\label{qdef}}
In this section, we describe the $q$-deformation of the Lie algebra $\suu$
into the algebra $\suq$ (also denoted $U_q(\suu)$),
% I don't like the triple close parens, but I like square brackets
% even less...  :-)
the representations of $\suq$, and the applications of $q$-deformation.
The deformations of $\spin(4)$ are then obtained through the isomorphism
$\spin(4)\isom\suu\oplus\suu$.

The following is part of the general subject of \emph{quantum groups}~\cite{qg-ref}.
Here we shall concentrate solely on the $\suu$ and $\spin(4)$ cases.

\subsection{The algebra $\suq$ and its representations}
The Lie algebra $\suu$ is generated by the well known Pauli matrices
$\sigma_i$, which obey the commutation relations
\begin{equation}\label{su2-lie}
	[\sigma_+,\sigma_-] = 4\sigma_3, \quad
	[\sigma_3,\sigma_+] = 2\sigma_+, \quad
	[\sigma_3,\sigma_-] =-2\sigma_-,
\end{equation}
where $\sigma_{\pm} = \sigma_1 \pm i\sigma_2$. The universal enveloping
algebra of $\suu$ is the associative algebra generated by $\sigma_\pm$
and $\sigma_3$ subject to the above identities, with the Lie bracket
being interpreted as $[A,B] = AB - BA$.

The $q$-deformed algebra $\suq$ is constructed by replacing $\sigma_3$
with another generator. Formally, it is thought of as $\Sigma =
q^{\frac{1}{2}\sigma_3}$, where $q\in\C$ with the exceptions $q\ne 0, 1,
-1$. The Lie bracket relations are replaced by the identities
\begin{equation}\label{su2q-ident}
	[\sigma_+,\sigma_-] = 4\frac{\Sigma^2-\Sigma^{-2}}{q-q^{-1}}, \quad
	\Sigma\sigma_+ = q\sigma_+ \Sigma, \quad
	\Sigma\sigma_- =-q\sigma_- \Sigma.
\end{equation}
We can rewrite $q=1+2\eps$ and think of $\eps$ as a small complex
number. Then, formally at leading order in $\eps$, the substitution
$\Sigma=q^{\frac{1}{2}\sigma_3}=1+\eps \sigma_3 + O(\eps^2)$ reduces the
deformed identities~\eqref{su2q-ident} to the standard Lie algebra
relations~\eqref{su2-lie}.  The associative algebra generated by
$\sigma_\pm$ and $\sigma_3$ subject to the deformed
identities~\eqref{su2q-ident} is the algebra $\suq$.

For generic $q$, that is, when $q$ is not a root of unity, the
finite-dimensional irreducible representations of $\suq$ are classified
by a half-integer, $j=0,1/2,1,3/2,\ldots,$ referred to as the
\emph{spin}, in direct analogy with the representations of $\suu$ and
the theory of angular momentum. The dimension of the representation $j$
is $2j+1$.  When $q=\exp(i\pi/r)$ is a $2r$th root of unity (ROU), where
$r>2$ is an integer called the ROU parameter, the representations $j$
are still defined, but become reducible for $j>(r-2)/2$. They decompose
into a sum of representations with spin at most $(r-2)/2$ and so-called
\emph{trace~0} ones, whose nature will be explained below.

For the purposes of this paper we are concerned only with intertwiners
between representations of $\suq$, i.e.,\ linear maps commuting with the
action of the algebra, and their (quantum) traces%
	\footnote{When $q=1$, this notion of trace reduces up to sign to the
	usual trace of a linear map, but is slightly
	different otherwise, cf.~\cite{CFS}*{Chapter~4}.}. %

Any such intertwiner can be constructed from a small set of generators
and elementary operations on them. These constructions, as well as
traces, can be represented graphically. Such graphs are called
\emph{(abstract) spin networks}.  Their calculus is well developed and
is described in~\cite{KL}, whose conventions we follow throughout the
paper with one exception: we use spins (half-integers) instead of
twice-spins (integers). A brief review of our notation and conventions
can be found in the Appendix.

Trace~0 representations of $\suq$ are so called because the trace of an
intertwiner from such a representation to itself is always zero. Thus,
they can be freely discarded, as they do not contribute to the
evaluation of $q$-deformed spin networks.

\subsection{Applications of $q$-deformation\label{qdef-appl}}
Deformation, especially with $q=\exp(i\pi/r)$ a $2r$th primitive ROU, is
important for spin foam models for at least two reasons. Replacing $q=1$
by some ROU can act as a regulator for a model whose partition function
and observable values are otherwise divergent.  Also, $\suq$ spin
networks%
	\footnote{These are graphs embedded in a $3$-manifold, labelled by
	representations of $\suq$. They are similar to but distinct from the
	abstract spin networks referred to above. See~\cite{Baez} for the
	distinction.} %
naturally appear when considering a positive cosmological constant in
loop quantum gravity.

The original Ponzano-Regge model~\cite{PR} attempts to express the path
integral for $3$-dimensional Riemannian general relativity as a sum over
labelled triangulations of a $3$-manifold. The edges of the
triangulation are labelled by discrete lengths, identified with spin
labels of irreducible $SU(2)$ representations. Each tetrahedron
contributes a $6j$-symbol factor to the summand, normalized to ensure
invariance of the overall sum under change of triangulation.
Unfortunately, the Ponzano-Regge model turned out to be divergent.
Motivated by the construction of $3$-manifold invariants, Turaev and
Viro were able to regularize the Ponzano-Regge model~\cites{TV,AW} by
replacing the $SU(2)$ $6j$-symbols with their $q$-deformed analogs at a
ROU $q$.  The key feature of the regularization is the truncation of the
summation to only the irreducible representations of $\suq$ of non-zero
trace, which leaves only a finite number of terms in the model's
partition function.

A version of the Barrett-Crane model, derived from a group field theory
by De~Pietri, Freidel, Krasnov and Rovelli~\cite{DFKR} (DFKR for short),
was also found to be divergent. A $q$-deformed version of the same model
at a ROU $q$ is similarly regularized (see Section~\ref{qdef-model}).
Some numerical results for the regularized version of this model are
given in Section~\ref{dfkr-reg}.

The argument linking $q$-deformation to the presence of a positive
cosmological constant is due to Smolin~\cite{Smolin} and  is given in
more refined form in~\cite{Smolin-posCC}.  It is briefly summarized as
follows. Loop quantum gravity begins by writing the degrees of freedom
of general relativity in terms of an $SU(2)$ connection on a spatial
slice and the slice's extrinsic curvature. A state in the Schr\"odinger
picture, a wave function on the space of connections, can be constructed
by integrating the Chern-Simons $3$-form over the spatial slice. This
state, known as the Kodama state, simultaneously satisfies all the
canonical constraints of the theory and semiclassically approximates
de~Sitter spacetime, which is a solution of the vacuum Einstein
equations with a positive cosmological constant.  The requirement that
the Kodama state also be invariant under large gauge transformations
implies discretization of the cosmological constant, $\Lambda \sim 1/r$,
with $r$ a positive integer.  The coefficients of the Kodama state in
the spin network basis are obtained by evaluating the labelled graph,
associated to a basis state, as an abstract $\suq$ spin network. Here
the deformation parameter $q$ is a ROU, $q=\exp(i\pi/r)$, where the ROU
parameter $r$ is identified with the discretization parameter of the
cosmological constant.

Given the heuristic link~\cite{Baez} between spin networks of loop
quantum gravity and spin foams, it is natural to $q$-deform a spin foam
model as an attempt to account for a positive cosmological constant.
With this aim, Noui and Roche~\cite{NR} have given a $q$-deformed
version of the Lorentzian Barrett-Crane model.  The possibility of
$q$-deformation has been with the Riemannian Barrett-Crane model since
its inception~\cite{BC-riem} and all the necessary ingredients have been
present in the literature for some time.  In the next section these
details are collected in a form ready for numerical investigation.

\section{Deformation of the Barrett-Crane model\label{bc-deform}}
Consider a triangulated $4$-manifold. Let $\Delta_n$ denote the set of
$n$-dimensional simplices of the triangulation. The dual $2$-skeleton is
formed by associating a dual vertex, edge and polygonal face to each
$4$-simplex, tetrahedron, and triangle of the triangulation,
respectively. A \emph{spin foam} is an assignment of labels, usually
called spins, to the dual faces of the dual $2$-skeleton. Each dual edge
has $4$ spins incident on it, while each dual vertex has $10$. A
\emph{spin foam model} assigns amplitudes $A_F$, $A_E$ and $A_V$, that
depend on all the incident spins, to each dual face, edge and vertex,
respectively. The amplitude $Z(F)$ assigned to a spin foam $F$ is the
product of the amplitudes for individual cells of the $2$-complex, while
the total amplitude $Z_\mathrm{tot}$ assigned to a triangulation is
obtained by summing over all spin foams based on the triangulation:
\begin{equation}\label{amp-partfunc}
	Z(F) = \prod_{f\in\Delta_2} A_F(f) \prod_{e\in\Delta_3} A_E(e)
		\prod_{v\in\Delta_4} A_V(v), \quad
	Z_\mathrm{tot} = \sum_{F} Z(F).
\end{equation}
Some models, such as those based on group field
theory~\cites{DFKR,Perez,GFT}, also include a sum over triangulations in
the definition of the total partition function.

\subsection{Review of the undeformed model}
The Riemannian Barrett-Crane model was first proposed in~\cite{BC-riem}.
Its relation to the Crane-Yetter~\cite{CY} spin foam model is analogous
to the relation of the Plebanski~\cite{Plebanski} formulation of general
relativity (GR) to $4$-dimensional $BF$ theory with $\Spin(4)$ as the
structure group. Both $BF$ theory and the Crane-Yetter model are
topological and the latter is considered a quantization of the
former~\cite{CY-BF}. In the Plebanski formulation, GR is a constrained
version of $BF$ theory. Similarly, the Barrett-Crane model restricts the
spin labels summed over in the Crane-Yetter model.  With this
restriction, Barrett and Crane hoped to produce a discrete model of
quantum (Riemannian) GR.

\subsubsection{Dual vertex amplitude}
All amplitudes are defined in terms of $\spin(4)$ spin networks.
However, given the isomorphism $\spin(4)\isom \suu\oplus\suu$, all
irreducible representations of $\spin(4)$ can be written as tensor
products of irreducible representations of $\suu$. The Barrett-Crane
model specifically limits itself to \emph{balanced} representations,
which are of the form $j\otimes j$, where $j$ is the irreducible
representation of $\suu$ of spin~$j$. Since the tensor product
corresponds to a juxtaposition of edges in a spin network, any
$\spin(4)$ spin network may be written as an $\suu$ spin network where
an edge labelled $j\otimes j$ is replaced by two parallel edges, each
labelled $j$. To avoid redundancy of notation, we use a single $j$
instead of $j\otimes j$ to label $\spin(4)$ spin network edges. We then
distinguish them from $\suu$ networks by placing a bold dot at every
vertex.

The Barrett-Crane vertex is an intertwiner between four balanced
representations:
\begin{equation}\label{BC-split}
	\bcv = \sum_{e} \frac{\bubj}{\thade \thbce}
		\hspace{.5em}\YYlow\hspace{.2em} \otimes
		\hspace{.5em}\YYlow\hspace{.2em}.
\end{equation}
The graphs on the right hand side of the definition are $\suu$ spin
networks and the sum runs over all admissible labels $e$. The graphical
notation and the conditions for admissibility are defined in the
Appendix.

The above expression defines the Barrett-Crane vertex in a way that
breaks rotational symmetry. However, it can be shown that the vertex is
in fact rotationally symmetric.  Up to normalization, this property
makes the Barrett-Crane vertex unique~\cite{BC-uniq}.  The above formula
defines a \emph{vertical splitting} of the vertex. A ninety degree
rotation will define an analogous \emph{horizontal splitting}. Both
possibilities are important in the derivation of the algorithm presented
in Section~\ref{num-10j}.

Given a $4$-simplex $v$ of a triangulation, the corresponding vertex of
the dual $2$-complex is assigned the amplitude
% TODO: Make the spin network diagrams slightly smaller.
\begin{equation}\label{10j-graph}
	A_V(v) = \;\;\tenjv\;\;.
\end{equation}
This spin network is called the \emph{$10j$-symbol}. The $4$-simplex $v$
is bounded by five tetrahedra, which correspond to the vertices of the
$10j$ graph. The four edges incident on a vertex correspond to the four
faces of the corresponding tetrahedron; the spin labels are assigned
accordingly. The edge joining two vertices corresponds to the face
shared by corresponding tetrahedra.  Evaluation of the $10j$-symbol is
discussed in Section~\ref{num-10j}. While the crossing structure
depicted above is immaterial in the undeformed case, it is essential at
nontrivial values of $q$. It is given here for reference.

\subsubsection{Dual edge and face amplitudes}
The original paper of Barrett and Crane did not specify dual edge and
face amplitudes. Three different dual edge and face amplitude
assignments were considered in a previous paper~\cite{BCHT}. We
concentrate on the same possibilities.

For the Perez-Rovelli model~\cite{PeRo}, we have
\begin{equation}
	A_F(f) = ~\vbubf, \qquad
	A_E(e) = \frac{\eye}{~~\vbubei~~\vbubeii~~\vbubeiii~~\vbubeiv}.
\end{equation}
For the DFKR model~\cite{DFKR}, we have
\begin{equation}
	A_F(f) = ~\vbubf, \qquad
	A_E(e) = \frac{1}{\eye}.
\end{equation}
For the Baez-Christensen model~\cite{BCHT}, we have
\begin{equation}
	A_F(f) = 1, \qquad
	A_E(e) = \frac{1}{\eye}.
\end{equation}

The bubble diagram, when translated into $\suu$ spin networks,
corresponds to two bubbles (see Appendix)
\begin{equation}\label{bub-eval}
	\vbubf
	= \left(\hspace{.25em}\bubj\right)^2.
\end{equation}
and evaluates to $(2j+1)^2$.

The so-called \emph{eye diagram} simply counts the dimension of the
space of $4$-valent intertwiners, which is also the number of admissible
$e$-edges summed over in Equation~\eqref{BC-split}. In symmetric form,
it is given by
\begin{equation}\label{eye-eval}
	\eye = \begin{cases}
		1+\min\{2j,s-2J\} & \text{if positive and $s$ is integral,} \\
		0 & \text{otherwise,}
	\end{cases}
\end{equation}
where $s=\sum_k j_k$, $j=\min_k j_k$, and $J=\max_k j_k$.

\subsection{The $q$-deformed model\label{qdef-model}}
Thanks to graphical notation, the $q$-deformation of the spin foam
amplitudes described above is straightforward, with only a few
subtleties. The main distinction is that $q$-deformed graphs are
actually ribbon (framed) graphs with braiding. Thus, any undeformed spin
network has to be supplemented with information about twists and
crossings before evaluation.

In~\cite{Yetter}, Yetter generalized the Barrett-Crane $4$-vertex for a
$q$-deformed version of $\spin(4)$.  Since $\spin(4)\isom
\suu\oplus\suu$, there is a two parameter family of possible
deformations of the Lie algebra, $\spin_{q,q'}(4) \isom \suq \oplus
\suqp$. Yetter singles out the one parameter family $q'=q^{-1}$,
restricted to balanced representations, since it preserves the
invariance of the Barrett-Crane vertex under rotations. This family also
has especially simple curl and twist identities:
\begin{equation}\label{curl-twist}
	 \curl \phantom{j} = \phantom{jj} \line \phantom{j\otimes}
		\quad \text{and} \quad
	\phantom{a\otimes} \tvertbt \phantom{b\otimes b}
		= \phantom{a\otimes a} \tvertt \phantom{b\otimes},
\end{equation}
where the left factor of $j\otimes j$ corresponds to $\suq$ and the
right one to $\suqi$, and the $3$-vertex is the obvious juxtaposition of
two $\suq$ and $\suqi$ $3$-vertices. Once this deformation is adopted,
the ribbon structure can be ignored~\cite{Yetter}, so one only needs to
specify the crossing structure for a given $\spin(4)$ spin network to
obtain a well-defined $q$-evaluation.

There are three basic graphs needed to define the Barrett-Crane simplex
amplitudes: the bubble, the eye, and the $10j$-symbol.  The evaluation
of the bubble graph, Equation~\eqref{bub-eval}, is $[2j+1]^2$, where the
quantum integer $[2j+1]$ is defined in the Appendix. Remarkably, the
value of the eye diagram turns out not to depend on $q$ and its value is
still given by Equation~\eqref{eye-eval}. The only exception is when $q$
is a ROU with parameter $r$. Then, the dimension of the space of
$4$-valent intertwiners changes to
\begin{equation}
	\eye = \begin{cases}
		\min\left\{\begin{matrix}
			\phantom{r-{}}1 + \min\{2j,s-2J\} \\
			r-1 - \max\{2J,s-2j\}
		\end{matrix}\right\} &
			\parbox{6.5em}{if positive and\\ $s$ is integral,} \\
		0 & \text{otherwise,}
	\end{cases}
\end{equation}
where again $s=\sum_k j_k$, $j = \min_k j_k$, and $J = \max_k j_k$.

The $10j$-symbol is the only network with a non-planar graph.
Originally, it was defined in terms of the $15j$-symbol from the
Crane-Yetter model. This $15j$-symbol was defined with $q$-deformation
in mind, so its crossing and ribbon structure was fully
specified~\cite{CKY}*{Section~3}. Adapted to the $10j$-graph, it can be
summarized as follows:
%% Alternative formulation
%\emph{Consider the complete graph on five vertices, $K_5$, embedded in
%the boundary of the $4$-simplex as the set of its edges and vertices.
%Remove a point from the interior of one of the tetrahedral faces of the
%$4$-simplex. The punctured boundary is now homeomorphic to $\R^3$. For
%any such homeomorphism, the embedding of $K_5$ into $\R^3$ can be
%projected onto a $2$-dimensional plane. The crossing structure of the
%$10j$ graph is defined by such a projection.}
%
\emph{Consider a $4$-simplex. The dual 1-skeleton of the boundary has
five dual vertices and ten dual edges, and is the complete graph $K_5$
on these five dual vertices.  If we remove one of the (non-dual)
vertices from the boundary of the $4$-simplex, what remains is
homeomorphic to $\R^3$.  For any such homeomorphism, the embedding of
$K_5$ into $\R^3$ can be projected onto a $2$-dimensional plane. The
crossing structure of the $10j$ graph is defined by such a projection.}
It is illustrated in Equation~\eqref{10j-graph}. Although, with
crossings, the $10j$ graph is no longer manifestly invariant under
permutations of its vertices, it can be shown to be so.

\subsection{Observables}\label{observables}
The definition of observables in a spin foam model of quantum gravity is
still open to interpretation (see Section~6 of~\cite{BCHT} for a brief
discussion). For a fixed spin foam, the half-integer spin labels of its
faces are the fundamental variables of the model. Practically speaking,
any observable of a spin foam model should be an expectation value of
some function $O(F)$ of the spin labels of a spin foam $F$, averaged
over all spin foams with amplitudes specified by
Equation~\eqref{amp-partfunc}:
\begin{equation}\label{obsv-sum}
	\ev{O} = \sum_{F} \frac{O(F) Z(F)}{Z_\mathrm{tot}}.
\end{equation}

In this paper we choose to concentrate on a few observables
representative of the kind of quantities computable in a spin foam
model. As before, fix a triangulation of a $4$-manifold, let $\Delta_2$
represent the set of its faces and let $j\colon \Delta_2 \to
\{0,1/2,1,\ldots\}$ be the spin labelling. We define:
\begin{align}
	J(F) &= \frac{1}{|\Delta_2|} \sum_{f\in\Delta_2}
		\qhi{j(f)}, \\
	(\delta J)^2(F) &= \frac{1}{|\Delta_2|} \sum_{f\in\Delta_2}
		\left(\qhi{j(f)}-\ev{J}\right)^2, \\
	A(F) &= \frac{1}{|\Delta_2|} \sum_{f\in\Delta_2}
		\sqrt{\qhi{j(f)}\qhi{j(f)+1}}, \\
        \label{eq:C_d}
	C_d(F) &= \frac{1}{N_d} \underset{\dist(f,f')=d}{\sum_{f, f' \in \Delta_2}}
		\frac{\qhi{j(f)}\qhi{j(f')} - \ev{J}^2}{\ev{(\delta J)^2}}.
\end{align}
where $\qhi{n}$ denotes a quantum half-integer (see Appendix), $|\cdot|$
denotes cardinality, $\dist(f,f')$ denotes the distance between faces,
and $N_d$ is a normalization factor (see below for the definition of
distance and $N_d$). These observables represent \emph{average spin per
face}, \emph{variance of spin per face}, \emph{average area per face},
and \emph{spin-spin correlation} as a function of $d$.

The choice of observables given above is somewhat arbitrary. For
instance, there are several subtly distinct choices for the expression
for $(\delta J)^2$. Fortunately, they all yield expectation values that
are nearly identical. The expression given above has the technical
advantage of falling into the class of so-called \emph{single spin
observables}. These are observables whose expectation value can be
directly obtained from the knowledge of probability with which spin $j$
occurs on any face of a spin foam. All of $J$, $(\delta J)^2$, and $A$
are single spin observables, while $C_d$ is not.

Note that on a fixed triangulation with no other background geometry,
there is no physical notion of distance. We can, instead, define a
combinatorial analog. For any two faces $f$ and $f'$ of a given
triangulation, let $\dist(f,f')$ be the smallest number of face-sharing
tetrahedra that connect $f$ to $f'$.  Given the discrete structure of
our spacetime model, it is conceivable that this combinatorial distance,
multiplied by a fundamental unit of length, approximates some notion of
distance derived from the dynamical geometry of the spin foam model.

The correlation function $C_d$ may be thought of as analogous to a
normalized $2$-point function of quantum field theory.  The
\emph{$d$-degree} of face $f$ is the number of faces $f'$ such that
$\dist(f,f') = d$.  If the $d$-degree of every face is the same, the
normalization factor $N_d$ can be taken to be the number of terms in the
sum~\eqref{eq:C_d}, that is, the number of face pairs separated by
distance $d$. This choice ensures the inequality $|C_d|\le 1$. If not
all faces have the same $d$-degree, then the normalization factor has to
be modified to
\begin{equation}
	N_d = |\Delta_2| D_d,
\end{equation}
where $D_d$ is the maximum $d$-degree of a face, which reduces to the
simpler definition in the case of uniform $d$-degree.

The choice of the  $q$-dependent expression $\qhi{j}$, instead of simply
using the half-integer $j$, is motivated in Section~\ref{rlim}. For some
$q$, the argument of the square root in $A(F)$ may be negative or even
complex. In that case, a branch choice will have to be made.  Luckily,
if $q=1$, $q$ is a ROU, or $q$ is real, the expression under the square
root is always non-negative.

\section{Numerical simulation\label{numsim}}
The key development that made possible numerical simulation of
variations of the (undeformed) Barrett-Crane model~\cites{BCHT,BCE} is
the development by Christensen and Egan of a fast algorithm for
evaluating $10j$-symbols~\cite{CE}. In this section, we show how this
algorithm generalizes to the $q$-deformed case and discuss numerical
evaluation of observables for the previously described spin foam models.

\subsection{The $q$-deformation of the fast $10j$ algorithm\label{num-10j}}
The derivation of the Christensen-Egan algorithm given in~\cite{CE} is
contingent on the possibility of splitting the Barrett-Crane $4$-vertex
as in Equation~\eqref{BC-split} and on the recoupling identity,
Equation~\eqref{rec-id} of the Appendix. Both identities still hold in
the $q$-deformed case. The validity of the $4$-vertex splitting was
proved by Yetter~\cite{Yetter} and the recoupling identity is a standard
part of $\suq$ representation theory.

\begin{figure}[tb]
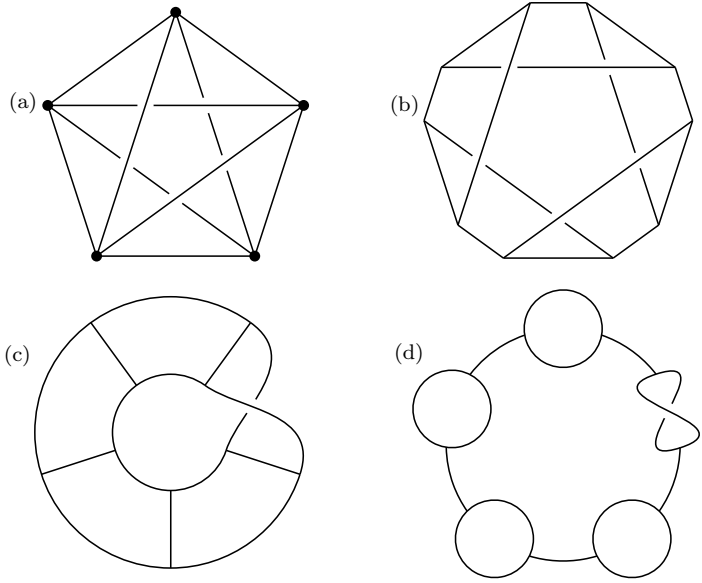

\centerline{%
	\begin{tabular}{cc}
		\tenj & \fifj \\
		\ladder & \hspace{2em}\bubbles
	\end{tabular}
}
	\caption{In reference to~\cite{CE}, (a) corresponds to Equation~(1),
	(b) corresponds to Equation~(2), while (c) and (d) correspond to the
	``ladder'' and ``bubble'' diagrams of Section~4, respectively. The
	illustrated twist introduces the explicitly $q$-dependent factor into
	Equation~\eqref{prefactor}. \label{bubble-twist}}
\end{figure}

The only remaining detail of the algorithm's generalization is the
crossing structure of the $10j$ graph, which was established in
Section~\ref{qdef-model}. However, its only consequence is an extra
factor from the twist implicit in the bubble diagram of Section~4 of
\cite{CE}, cf.\ Equation~\eqref{qtwist} of the Appendix. We will not
reproduce the derivation of the algorithm here. However, the way in
which the twist arises is schematically illustrated in
Figure~\ref{bubble-twist}. Note that the triviality of the twist for
Yetter's balanced representations, Equation~\eqref{curl-twist}, does not
apply here since the twist occurs separately in distinct $\suq$
networks.

The algorithm itself can be summarized in the following form:
\begin{equation}
	\{10j\} = (-)^{2S} \sum_{m_1,m_2} \phi \tr[M_4 M_3 M_2 M_1 M_0].
\end{equation}
The $10j$-symbol depends on the ten spins $j_{i,k}$, ($i=1,2$,
$k=0,\ldots,4$) specified in Equation~\eqref{10j-graph}. The overall
prefactor depends on the total spin $S = \sum_{i,k} j_{i,k}$
and the per-term prefactor is
\begin{equation}\label{prefactor}
	\phi = (-)^{m_1-m_2} [2m_1+1] [2m_2+1] q^{m_1(m_1+1)-m_2(m_2+1)}.
\end{equation}
The exponents of $(-)$ and $q$ are always integers. The $M_k$ are
matrices (not all of the same size) of dimensions compatible with the
five-fold product and trace. Their matrix elements are
\begin{align}
	(M_k)_{l_k}^{l_{k+1}} &=
	\frac{[2l_k+1] (T_1)_{l_k}^{l_{k+1}} (T_2)_{l_k}^{l_{k+1}}}
		{\theta(j_{2,k-1},l_{k+1},j_{1,k})\,\theta(j_{2,k+1},l_{k+1},j_{1,k+1})}, \\
	(T_i)_{l_k}^{l_{k+1}} &=
	\frac{\Tet{ l_k & j_{2,k} & m_i \cr l_{k+1} & j_{2,k-1} & j_{1,k}}}
		{\theta(j_{2,k},l_{k+1},m_i)}.
\end{align}
The quantum integers $[n]$, as well as the theta $\theta(a,b,c)$ and
tetrahedral $\operatorname{Tet}[{\cdots}]$ $\suq$ spin networks
are defined in the Appendix.

The quantities $l_k$ and $m_i$ are spin labels (half-integers). They are
constrained by admissibility conditions (parity conditions and triangle
inequalities).  The parity of each index is determined by the conditions
\begin{align}
\label{l-par}
	l_k &\equiv j_{1,k} + j_{2,k} \equiv j_{1,k-1} + j_{2,k-2}, \\
\label{m-par}
	m_i &\equiv l_k + j_{2,k-1},
\end{align}
for $i=1,2$ and $k=0,\ldots,4$, where $\equiv$ denotes equivalence
mod~$1$ and the second subscript of $j$ is taken mod~$5$.  Summation
bounds are determined by the triangle inequalities, which must be
checked for each trivalent vertex introduced in the derivation of the
algorithm. They boil down to
\begin{align}
	\lb_3(j_{1,k},j_{2,k},j_{2,k-1}) &\le m_i \le j_{1,k}+j_{2,k}+j_{2,k-1}, \\
	|j_{1,k-1}-j_{2,k-2}| &\le l_k \le j_{1,k-1}+j_{2,k-2}, \\
	|j_{1,k}-j_{2,k}| &\le l_k \le j_{1,k}+j_{2,k}, \\
	|m_i-j_{2,k-1}| &\le l_k \le m_i+j_{2,k-1},
\end{align}
for $i=1,2$ and $k=0,\ldots,4$, where we have used the notation
\begin{align}
	\lb_3(a,b,c) &= 2\max\{a,b,c\}-(a+b+c).
\end{align}
When $q=\exp(i\pi/r)$ is a ROU, extra inequalities must be taken into
account to exclude summation over reducible representations.  These are
\begin{align}
	m_i &\ge j_{1,k}+j_{2,k}+j_{2,k-1} - (r-2), \\
	m_i &\le \ub_3(j_{1,k},j_{2,k},j_{2,k-1}) + (r-2), \\
	l_k &\le (r-2)-(j_{1,k}+j_{2,k}), \\
	l_k &\le (r-2)-(j_{1,k-1}+j_{2,k-2}), \\
	l_k &\le (r-2)-(m+j_{2,k-1}),
\end{align}
where now
\[
	\ub_3(a,b,c) = 2\min\{a,b,c\}-(a+b+c).
\]
If any of the parity constraints or inequalities cannot be satisfied,
the $10j$-symbol evaluates to zero.

This algorithm has been implemented and tested in the $q=1$ and ROU
cases, for both $j$ and $r$ up to several hundreds. Unfortunately, for
generic $q$, when $Q=\max\{|q|,|q|^{-1}\} > 1$, the quantum integers
grow exponentially as $|[n]| \sim Q^n$. Such a rapid growth makes the
sums involved in this algorithm numerically unstable. It is still
possible to use this algorithm with $Q$ close to $1$ or symbolically,
using rational functions of $q$ instead of limited precision floating
point numbers.  Symbolic computation is, however, significantly slower
(by up to a factor of $10^6$) than its floating point counterpart.  The
software library \texttt{spinnet} which implements these and other spin
network evaluations is available from the authors and will be described
in a future publication.

\subsection{Positivity and statistical methods}
The sums involved in evaluating expectation values of observables, as in
Equation~\eqref{obsv-sum}, are very high-dimensional. For instance, a
minimal triangulation of the $4$-sphere (seen as the boundary of a
$5$-simplex) contains $20$ faces.  Hence, any brute force evaluation of
an expectation value, even on such a small lattice, involves a sum over
the $20$-dimensional space of half-integer spin labels.

Fortunately, in the undeformed case, the total amplitude $Z(F)$ for a
closed spin foam is never negative%
	\footnote{We expect the same thing to hold in Lorentzian
	signature~\cites{BC-pos,CC}.}%
~\cite{BC-pos}. The proof for the $q=1$ case generalizes to the ROU
case. One need only realize two facts. The first is that, in the ROU
case, quantum integers are non-negative. The second is that, for $q$ a
ROU, an $\suqi$ spin network evaluates to the complex conjugate of the
corresponding $\suq$ spin network.  The disjoint union of any two such
spin networks evaluates to their product, the absolute value squared of
either of them, and hence is non-negative. Then, the same positivity
result follows as from Equation~(1) of~\cite{BC-pos}.  This positivity
allows us to treat $Z(F)$ as a statistical distribution and use Monte
Carlo methods to extract expectation values with much greater efficiency
than brute force summation.

The main tool for evaluating expectation values is the Metropolis
algorithm~\cites{Metrop,LB}. The algorithm consists of a walk on the
space of spin labellings. Each step is randomly picked from a set of
\emph{elementary moves} and is either accepted or rejected based on the
relative amplitudes of spin foam configurations before and after the
move. An expectation value is extracted as the average of the observable
over the configurations constituting the walk. Elementary moves for spin
foam simulations are discussed in the next section.

A Metropolis-like algorithm is possible even if individual spin foam
amplitudes $Z(F)$ are negative or even complex. However, if the total
partition function $Z_\mathrm{tot}$ sums to zero, then the expectation
values in Equation~\eqref{obsv-sum} become ill defined. Moreover, in
numerical simulations, if $Z_\mathrm{tot}$ is even close to zero,
expectation value estimates may exhibit great loss of precision and slow
convergence. In the path-integral Monte Carlo literature, this situation
is known as the \emph{sign problem}~\cite{sgn-prob}. Still, the sign
problem need not occur or, depending on the severity of the problem,
there may be ways of effectively dealing with it.

Independent Metropolis runs can be thought of as providing independent
estimates of a given expectation value. Thus, the error in the computed
value of an observable can be estimated through the standard deviation
of the results of many independent simulation runs~\cite{KI}.

\subsection{Elementary moves for spin foams}
The choice of elementary moves for spin foam simulations must satisfy
several criteria. Theoretically, the most important one is ergodicity.
That is, any spin foam must be able to transform into any other one
through a sequence of elementary moves which avoid configurations with
zero amplitude.  Practically, it is important that these moves usually
preserve admissibility. A spin foam $F$ is called \emph{admissible} if
the associated amplitude $Z(F)$ is non-zero. If, starting with an
admissible spin foam, most elementary moves produce an inadmissible spin
foam, the simulation will spend a lot of time rejecting such moves
without any practical benefit.

As before, consider a fixed triangulation of a compact $4$-manifold.
The parity conditions~\eqref{l-par} imposed on the $j_{i,k}$,
\[
  j_{1,k} + j_{2,k} \equiv j_{1,k-1} + j_{2,k-2}, \quad 0 \leq k \leq 4,
\]
when taken together with the total spin foam
amplitude~\eqref{amp-partfunc}, provide strong constraints on admissible
spin foams.  One can show that a move that changes spin labels by $\pm
1/2\pmod{1}$ on each face of a closed surface in the dual $2$-skeleton
preserves the parity constraint.  We take as the elementary moves the
moves that change the spin labels by $\pm 1/2$ on the boundaries of the
dual $3$-cells of the dual $3$-complex; the dual $3$-cells correspond to
the edges of the triangulation.  If the manifold has non-trivial mod~$2$
homology in dimension $2$, additional moves would be necessary, but for
the examples we consider the moves above suffice. From a practical point
of view, extra moves might improve the simulation's equilibration time.
For instance, in the ROU case, parity preserving moves that change the
spins from $0$ to $(r-2)/2$ or $(r-3)/2$ were introduced, since spins
close to either admissible extreme may have large amplitudes. This
property of the Perez-Rovelli and Baez-Christensen models is illustrated
in the following section.

Unfortunately, the inequalities constraining spin labels do not have a
similar geometric interpretation and cannot be used to easily restrict
the set of elementary moves in advance.

\section{Results\label{results}}
Using methods described in the previous section, we ran simulations of
the three variations of the Barrett-Crane model described in
Section~\ref{bc-deform} and obtained expectation values for observables
listed in Section~\ref{observables}. While previous work~\cite{BCHT}
performed simulations only on the minimal triangulation of the
$4$-sphere, which we will refer to simply as the \emph{minimal
triangulation}, we have extended the same techniques to arbitrary
triangulations of closed manifolds.

\subsection{Discontinuity of the $r\to\oo$ limit\label{rlim}}
The most striking result we can report is a discontinuity in the
transition to the limit $r\to\oo$, where $r$, a positive integer, is the
ROU parameter with $q=\exp(i\pi/r)$. As $r\to\oo$, the deformation
parameter $q$ tends to its classical value $1$. If we interpret the
cosmological constant as inversely proportional to $r$, $\Lambda \sim
1/r$, this limit also corresponds to $\Lambda \to 0$, through positive
values. For a fixed spin foam, the amplitudes and observables we study
tend continuously to their undeformed values as $r \to \oo$. However, we
find that observable \emph{expectation values} do not tend to their
undeformed values in the same limit, that is, $\ev{O}_r \nto
\ev{O}_{q=1}$ as $r\to\oo$.

\begin{figure}
	\centerline{%
	\hspace*{0.3cm}
	\includegraphics[width=7.1cm]{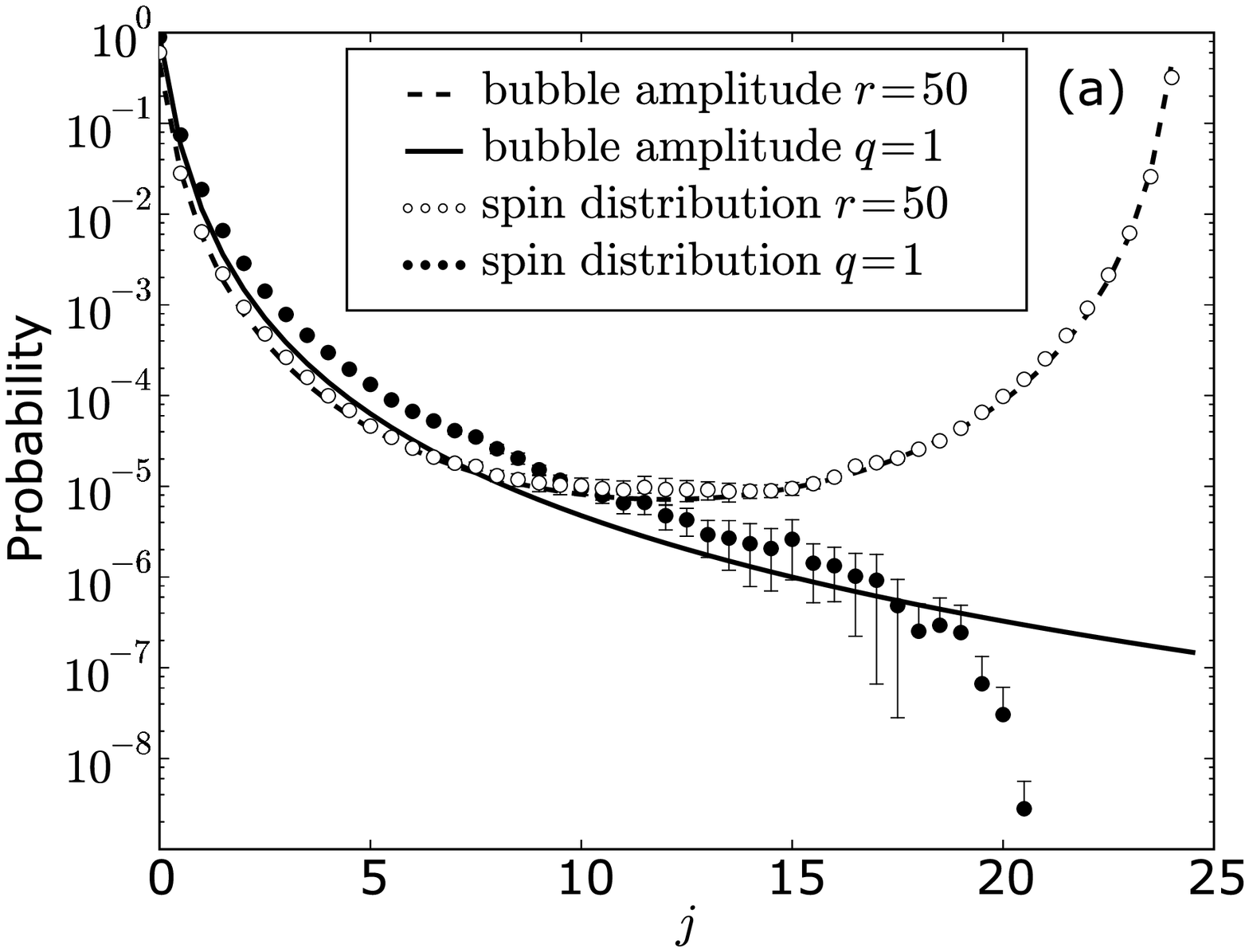}
	\hspace*{-0.7cm}
	\includegraphics[width=7.1cm]{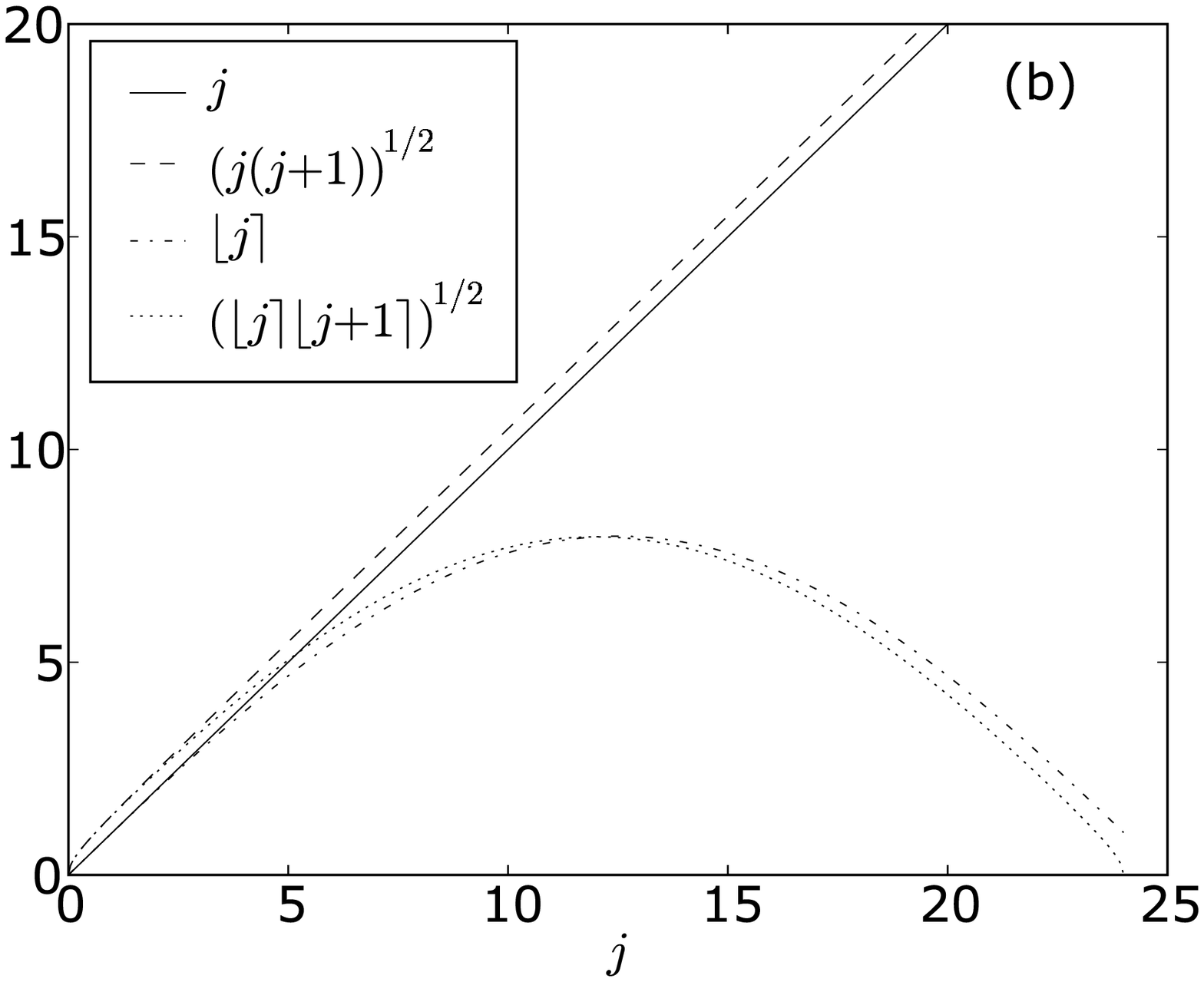}}
	\caption{(a) Single spin distribution and single bubble amplitude for
	the Baez-Christensen model. The distribution was obtained from 
	$10^{9}$ steps of Metropolis simulation on a triangulation with $202$
	faces (cf.\ Section~\ref{spincor}). (b) Some single spin observables as
	functions of $j$, with $r=50$.\label{spin-dist-bc}}
\end{figure}

The discontinuity is most simply illustrated with the \emph{single spin
distribution}, that is the probability of finding spin $j$ at any spin
foam face. This probability can be estimated from the histogram of all
spin labels that have occurred during a Monte Carlo simulation. The
points in Figure~\ref{spin-dist-bc}(a) 
% TODO: Shrink the white borders for these graphs, get them bigger and closer.
show the single spin distributions for the Baez-Christensen model with
$r=50$ and $q=1$. The curves show the corresponding \emph{single bubble}
amplitude.  It is the amplitude $Z(F_j)$ of a spin foam $F_j$ with all
spin labels zero, except for the boundary of an elementary dual
$3$-cell, whose faces are all labelled with spin $j$. The amplitudes and
distributions are normalized as probability distributions so their sums
over $j$ yield $1$.  The similarity between the points and the
continuous curves is consistent with the hypothesis that spin foams with
isolated bubbles dominate the partition function sum. The behavior of
the single spin distribution for the Perez-Rovelli model is very
similar, except that its peaks are much more pronounced.

Note that the undeformed single spin distribution has a single peak at
$j=0$, while the $r=50$ case has two peaks, one at $j=0$ and the other
at $j=(r-2)/2$, the largest non-trace~0 irreducible representation.  The
bimodal nature of the single spin distribution has an important impact
on the large $r$ behavior of observable expectation values, as is most
easily seen with single spin observables (Section~\ref{observables}).
For instance, if we consider the average, $\bar{j}$, of the
half-integers $j$, the large $j$ peak would dominate the expectation
value and $\ev{\bar{j}}$ would diverge linearly in $r$, as $r\to\oo$. On
the other hand, since $J$ is the average of the quantum half-integers
$\qhi{j}$, $\ev{J}$ at least approaches a constant in the same limit.
This is illustrated in Figure~\ref{spin-dist-bc}(b).

\begin{figure}[tb]
	\centerline{\includegraphics[width=10cm]{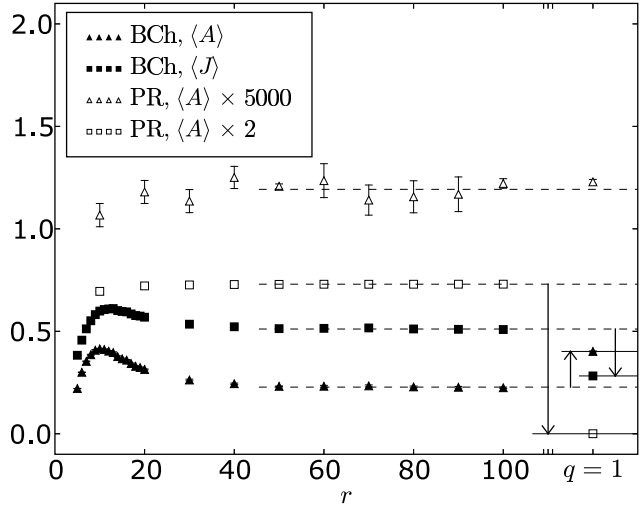}}
	\caption{Observables for the Baez-Christensen (BCh) and Perez-Rovelli
	(PR) models as functions of the ROU parameter $r$. For large $r$,
	observables do not in general tend to their undeformed, $q=1$, values;
	arrows show the deviation.
	Some observables were scaled to fit on the graph. Data is from Metropolis
	simulations on the minimal triangulation.\label{prbc-obsv}}
\end{figure}
However, as shown in Figure~\ref{prbc-obsv}, this limit is not the same
as the undeformed expectation value. At the same time, as can be seen
from the plot of the Perez-Rovelli average area in the same figure,
there are some observables whose large $r$ limits are at least very
close to the undeformed values. The area observable summand $A_j =
\sqrt{\qhi{j}\qhi{j+1}}$ is exactly zero at both $j=0$ and $j=(r-2)/2$,
while the spin observable summand $J_j=\qhi{j}$ is zero at $j=0$ but
still positive at $j=(r-2)/2$, Figure~\ref{spin-dist-bc}(b). The large
$j$ peak of the Perez-Rovelli model is very narrow and thus the
expectation value of a single spin observable is strongly influenced by
its value at $j=(r-2)/2$.

The data for larger triangulations is qualitatively similar.

\subsection{Regularization of the DFKR model\label{dfkr-reg}}
\begin{figure}[tb]
	\centerline{\includegraphics[width=10cm]{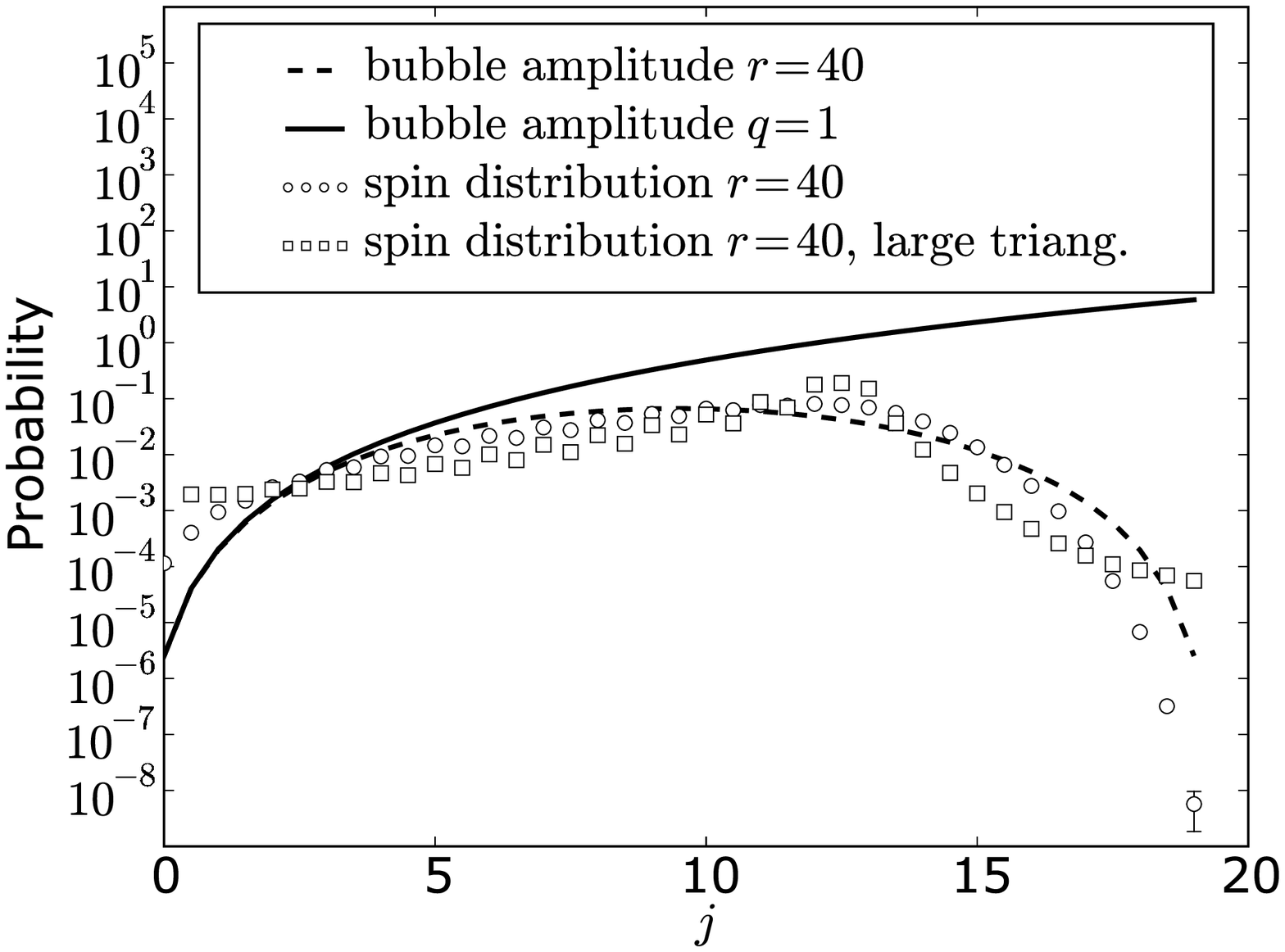}}
	\caption{Single spin distributions and single bubble amplitudes for the
	DFKR model. The distributions were obtained from 
	$10^9$ steps of Metropolis simulation on the minimal triangulation and
	on a triangulation with $202$
	faces (cf.\ Section~\ref{spincor}).\label{spin-dist-dfkr}}
\end{figure}
As expected, the ROU deformation of the DFKR model yields a finite
partition function and finite expectation values. For instance, its
single spin distribution for $r=40$ is illustrated in
Figure~\ref{spin-dist-dfkr}. The divergence of the amplitude for large
spins in the undeformed, $q=1$, case makes numerical simulation
impossible without an artificial spin cutoff. Thus, we do not have an
undeformed analog of the single spin distribution. For the minimal
triangulation, the ROU spin distribution deviates slightly from the
single bubble amplitude close to the boundaries of admissible $j$. For
the larger triangulation, the deviation is much more pronounced and is
not restricted to the edges.  This suggests that there are other
significant contributions to the partition function besides single
bubble spin foams.

\begin{figure}[tb]
	\centerline{\includegraphics[width=11cm]{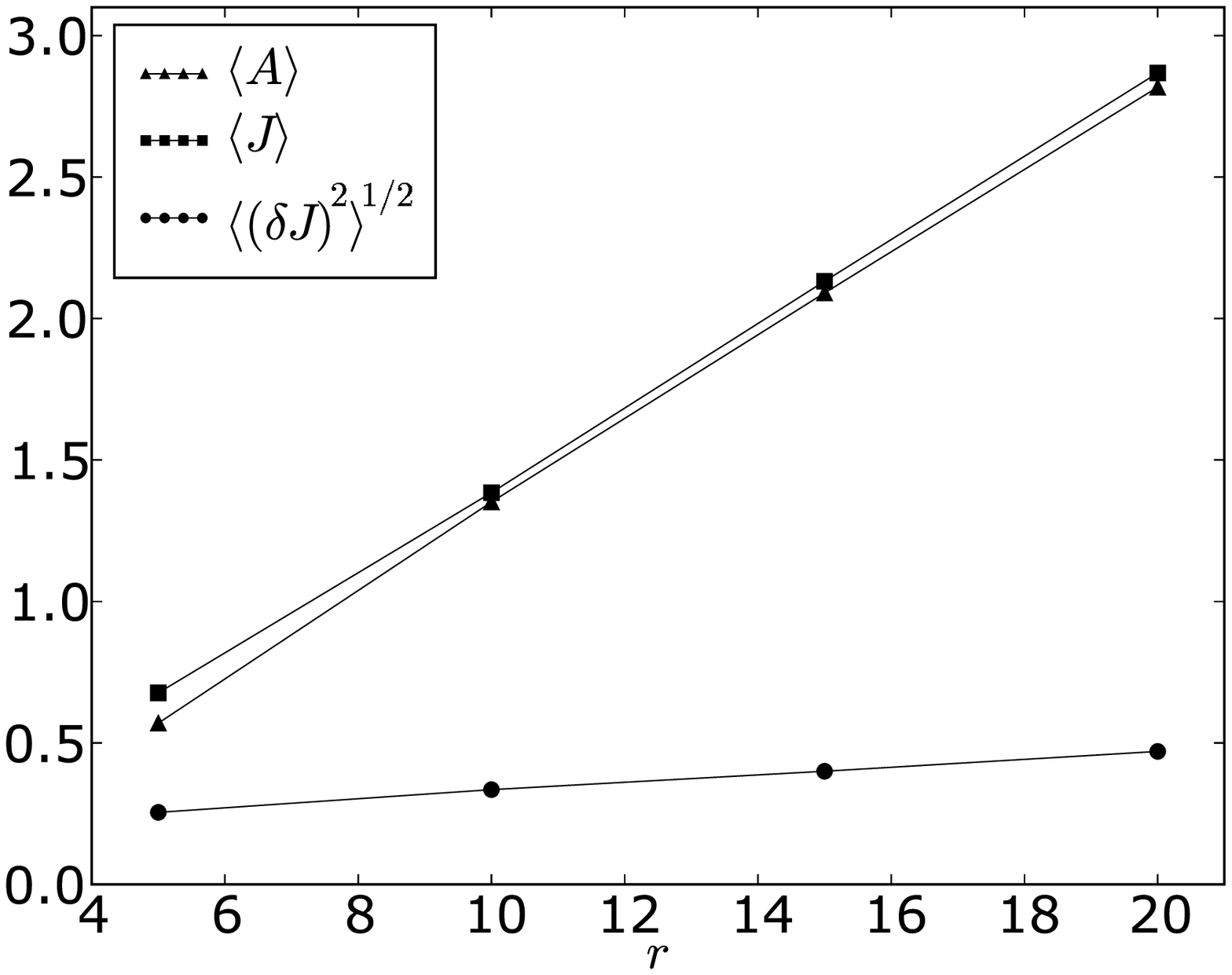}}
	\caption{Observables for the DFKR model: area $\ev{A}$, average spin
	$\ev{J}$, spin standard deviation $\sqrt{\ev{(\delta J)^2}}$.
	Metropolis simulation, minimal triangulation.
	Error bars are smaller than the data points.\label{dfkr-obsv}}
\end{figure}

Note the large weight associated with spins around $j=r/4$. Around this
value of $j$, both the area $A_j=\sqrt{\qhi{j}\qhi{j+1}}$ and the spin
$J_j=\qhi{j}$ attain their maximal values and are proportional to $r$.
Thus, it is natural to expect their expectation values to grow linearly
in $r$, which is consistent with the divergent nature of the undeformed
DFKR model. This is precisely the behavior shown in
Figure~\ref{dfkr-obsv}. On the minimal triangulation, the best linear
fits for the average spin expectation value and for the square root of
the average spin variance are
\begin{align}
	\ev{J}_r &= 0.146\, r - 0.064, \\
	\ev{(\delta J)^2}_r^{1/2} &= 0.014\, r + 0.187.
\end{align}
For larger triangulations, the dependence of these observables is also
approximately linear in $r$, with only slight variation in the effective
slope.

\subsection{Spin-spin correlation\label{spincor}}
The ability to work with larger lattices allows us to explore a broader
range of observables. One of them is the spin-spin correlation function
$C_d$ defined in Section~\ref{observables}. In general $\ev{C_0} = 1$
and $\ev{C_d}\to0$ for large $d$. The decay of the correlation shows how
quickly the spin labels on different spin foam faces become independent.
A positive value of $\ev{C_d}$ indicates that, on average, any two faces
distance $d$ apart both have spins above (or both below) the mean
$\ev{J}$. On the other hand, a negative value of $\ev{C_d}$ indicates
that, on average, any two faces distance $d$ apart have one spin above
and one below the mean $\ev{J}$.

% tb means to allow floats at tops and bottoms of pages, but
% not on a dedicated ``float'' page.  tbf is the default.
\begin{figure}[tb]
	\centerline{\includegraphics[width=11cm]{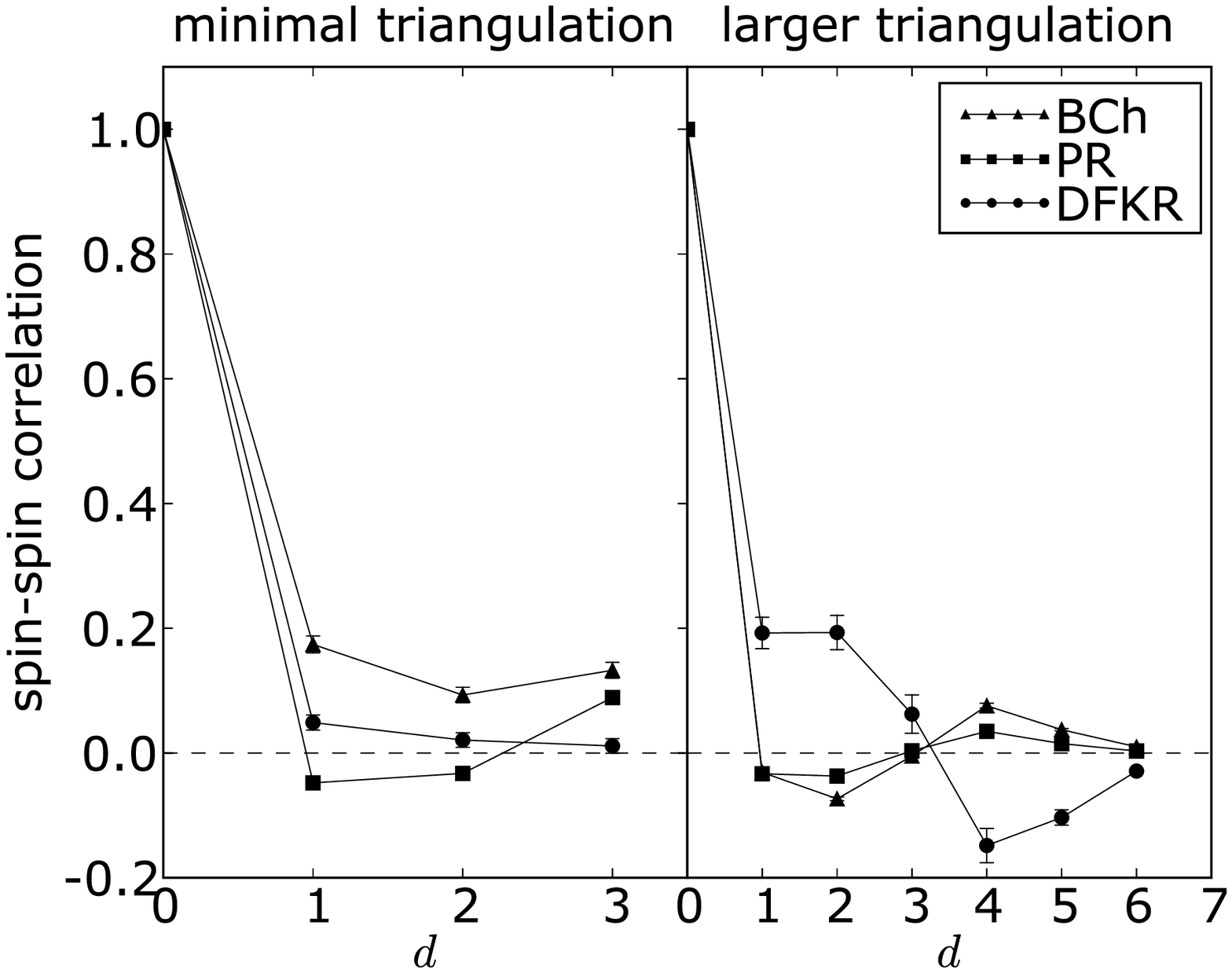}}
	\caption{Spin-spin correlation functions for the Baez-Christensen
	(BCh), Perez-Rovelli (PR) and DFKR models, on the minimal
	triangulation ($6$ vertices, $15$ edges, $20$ faces, $15$ tetrahedra,
	and $6$ $4$-simplices) as well as a larger triangulation ($23$
	vertices, $103$ edges, $202$ faces, $200$ tetrahedra, and $80$
	4-simplices). ROU parameter $r=10$. \label{plot-cor}}
\end{figure}

A small triangulation limits the maximum distance between faces. For
example, the minimal triangulation has maximum distance $d=3$.  Larger
triangulations of the $4$-sphere were obtained by refining the minimal
one by applying Pachner moves randomly and uniformly over the whole
triangulation.  We restricted the Pachner moves to those that did not
decrease the number of simplices.

The largest triangulation we have used has maximum distance $d=6$. Its
correlations for different models are shown in Figure~\ref{plot-cor}
along with those from the minimal triangulation. Correlation functions
for different values of ROU parameter $r$ (including the $q=1$ case) and
other triangulations are qualitatively similar.

Notice the small negative dip for small values of $d$ for the
Perez-Rovelli and Baez-Christensen models. As discussed in previous
sections, the partition functions of these models are dominated by spin
foams with isolated bubbles. The correlation data is consistent with
this hypothesis. The values of the spins assigned to faces of the bubble
will be strongly correlated, while the values of the spins on two faces,
one of which lies on the bubble and the other does not, should be
strongly anti-correlated. Since a given face usually has fewer nearest
neighbors that lie on the same bubble than that do not, on average, the
short distance correlation is expected to be negative. At slightly
larger distances, the correlation function turns positive again. This
indicates that on a larger triangulations, spin foams with several
isolated bubbles contribute strongly to the partition function.
Although, with so few data points, it is difficult to extrapolate the
behavior of the correlation function to larger triangulations and
distances, its features are qualitatively similar to that of a condensed
fluid, where the density-density correlation function exhibits
oscillations on the scale of the molecular dimensions.

Note that the behavior of the DFKR correlation function is significantly
different from the other two. This is also consistent with the already
observed fact that its partition function has strong contributions from
other than single or isolated bubble spin foams.

\section{Conclusion}\label{se:conclusion}
We have numerically investigated the behavior of physical observables
for the Perez-Rovelli, DFKR, and Baez-Christensen versions of the
Barrett-Crane spin foam model. Each version assigns different dual edge
and face amplitudes to a spin foam, and these choices greatly affect the
behavior of the resulting model. The behavior of the models was also
greatly affected by $q$-deformation.

The limiting behavior of observables was found to be discontinuous in
the limit of large ROU parameter $r$, i.e., $q=\exp(i\pi/r)$ close to
its undeformed value of $1$. This result is at odds with the physical
interpretation of the relation $\Lambda \sim 1/r$ between the
cosmological constant $\Lambda$ and the ROU parameter. Finally, the
behavior of the examined physical observables, especially of the
spin-spin correlation function, indicates the dominance of isolated
bubble spin foams in the Perez-Rovelli and Baez-Christensen partition
functions, while less so for the the DFKR one.

Some questions raised by these results deserve attention. For instance,
it is not known whether the same $q\to 1$ limit behavior will be
observed when $q$ is taken through non-ROU values. While calculations
with $\max\{|q|,|q|^{-1}\}>1$ are numerically unstable, they should
still be possible for $|q|\sim 1$. 

Another important project is to perform a more extensive study of the
effects of triangulation size in order to better understand the
semi-classical limit. 

Finally, all of this work should also be carried out for the Lorentzian
models, which are physically much more interesting but computationally
much more difficult.

These and other questions will be the subject of future investigations.

\section*{Acknowledgements}
The authors would like to thank Wade Cherrington for helpful
discussions. The first author was supported by NSERC and FQRNT
postgraduate scholarships and the second author by an NSERC grant.
Computational resources for this project were provided by SHARCNET.

\appendix
\section{Spin network notation and conventions}
\emph{Quantum integers} are a $q$-deformation of integers. For an integer $n$,
the corresponding quantum integer is denoted by $[n]$ and is given by
\begin{equation}
	[n] = \frac{q^n-q^{-n}}{q-q^{-1}}.
\end{equation}
In the limit $q\to 1$, we recover the regular integers, $[n]\to n$.
Note that $[n]$ is invariant under the transformation $q\mapsto q^{-1}$.
When $q=\exp(i\pi/r)$ is a root of unity (ROU), for some integer $r>1$,
an equivalent definition is
\begin{equation}
	[n] = \frac{\sin(n\pi/r)}{\sin(\pi/r)}.
\end{equation}
This expression is non-negative in the range $0\le n \le r$.
\emph{Quantum factorials} are defined as
\begin{equation}
	[n]! = [1][2]\cdots[n].
\end{equation}
In many cases, $q$-deformed spin network evaluations can be obtained
from their undeformed counterparts by simply replacing factorials with
quantum factorials. For convenience, when dealing with half-integral
spins, we also define \emph{quantum half-integers} as
\begin{equation}
	\qhi{j} = \frac{[2j]}{2}
\end{equation}
when $j$ is a half-integer.

Abstract $\suq$ spin networks can be approached from two different
directions. They can represent contractions and compositions of
$\suq$-invariant tensors and intertwiners~\cite{CFS}. At the same time,
they can represent traces of tangles evaluated according to the rules of
the Kauffman bracket~\cite{KL}. Either way, the computations turn out to
be the same. We present here formulas for the evaluation of a few spin
networks of interest.

The \emph{single bubble} network evaluates to what is sometimes called
the \emph{superdimension} of the spin-$j$ representation:
\begin{equation}
	\bubj = (-)^{2j} [2j+1].
\end{equation}
(As in the rest of the paper, the spin labels are half-integers.)

Up to a constant, there is a unique $3$-valent vertex (corresponding to
the Clebsch-Gordan intertwiner) whose normalization is fixed up to sign
by the value of the \emph{$\theta$-network}:
\begin{equation}
	\theta(a,b,c) = \thabc =
		\frac{(-)^s [s+1]! [s-2a]! [s-2b]! [s-2c]!}{[2a]! [2b]! [2c]!},
\end{equation}
where $s = a+b+c$. The $\theta$-network is non-vanishing, together with
the three-vertex itself, if and only if $s$ is an integer and the
triangle inequalities are satisfied: $a\le b+c$, $b\le c+a$, and $c\le
a+b$. In addition, when $q$ is a ROU, one extra inequality must be
satisfied: $s\le r-2$. The triple $(a,b,c)$ of spin labels is called
\emph{admissible} if $\theta(a,b,c)$ is non-zero.

The recoupling identity gives the transformation between different bases
for the linear space of $4$-valent tangles (or intertwiners):
\begin{equation}\label{rec-id}
	\YYh = \sum_{e} \frac{(-)^{2e} [2e+1]
			\Tet{ a & b & e \\ c & d & f }}{\theta(a,d,e)\,\theta(c,b,e)}
			\hspace{.5em}\YY,
\end{equation}
where the sum is over all admissible labels $e$ and the value of the
\emph{tetrahedral network} is
\begin{equation}
	 \Tet{ a & b & e \\ c & d & f } = \tet
	 = \frac{\mathcal{I}!}{\mathcal{E}!}
		\sum_{m\le S\le M} \frac{(-)^S [S+1]!}{\prod_i[S-a_i]!
			\prod_j[b_j-S]!},
\end{equation}
where
\begin{align}
	\mathcal{I}! &= \prod_{i,j}[b_j-a_i]! &
		\mathcal{E}! &= [2A]![2B]![2C]![2D]![2E]![2F]! \\
	a_1 &= (a+d+e) & b_1 &= (b+d+e+f) \\
	a_2 &= (b+c+e) & b_2 &= (a+c+e+f) \\
	a_3 &= (a+b+f) & b_3 &= (a+b+c+d) \\
	a_4 &= (c+d+f) & m &= \max\{a_i\} \quad M = \min\{b_j\}.
\end{align}
Due to parity constraints, the $a_i$, $b_j$, $m$, $M$, and $S$ are all
integers.

Since the three-vertex is unique up to scale, its composition with with
a braiding applied to two incoming legs yields a multiplicative factor:
\begin{equation}\label{qtwist}
	\tvertb
	= (-)^{a+b-c} q^{a(a+1)+b(b+1)-c(c+1)}
	\tvert.
\end{equation}
Note that the above braiding factor is not invariant under the
transformation $q\mapsto q^{-1}$, while the bubble, tetrahedral and
$\theta$-networks are all invariant under this transformation, by virtue
of their expressions in terms of quantum integers.


\begin{thebibliography}{99}
%%% Format amsrefs bibliography to IOP standards.
\makeatletter%
	\newcommand{\oparen}[1]{\upn{(}#1}%
	\newcommand{\cparen}[1]{\leavevmode\push@bracket)#1\pop@bracket}%
\makeatother%
\newcommand{\itcaps}[1]{\textit{\EnglishInitialCaps{#1}}}%
\newcommand{\eparens}[1]{\parenthesize{\textit{Preprint} \eprint{#1}}}%
\newcommand{\PrintEditorsIOP}[1]{\PrintNames{ed}{}{}{#1}}%
\renewcommand{\editiontext}{ed}%
\renewcommand{\etaltext}{\textit{et al}}%
\renewcommand{\sameauthors}[1]{\PrintNames{author}{}{}{#1}}%
% Author name formatting.
\def\InitialPunct{ }
\BibSpec{firstauthor}{%
	+{}{}{surname}
	+{}{ }{initials}
	+{}{ }{jr}
}%
\BibSpec{coauthor}{%
	+{}{ and}{transition}
	+{}{ }{surname}
	+{}{ }{initials}
	+{}{ }{jr}
}%
\BibSpec{middleauthor}{%
	+{,}{}{transition}
	+{}{ }{surname}
	+{}{ }{initials}
	+{}{ }{jr}
}%
\BibSpec{lastauthor}{%
	+{,}{ and}{transition}
	+{}{ }{surname}
	+{}{ }{initials}
	+{}{ }{jr}
}%
% Editor name formatting.
\BibSpec{middleed}{}%
\BibSpec{lasted}{%
	+{}{ \SubEtal}{surname}
}%
% Reference formatting.
\BibSpec{article}{%
	+{}{\PrintAuthors}{author}
	+{}{ \PrintDateB} {date}
	+{}{ \textit}     {journal}
	+{}{ \textbf}     {volume}
	+{}{ \eprintpages}{pages}
	+{}{ \eparens}    {eprint}
}%
\BibSpec{eprint}{%
	+{}{\PrintAuthors}             {author}
	+{}{ \PrintDateB}              {date}
	+{}{ }                         {title}
	+{}{ \textit{Preprint} \eprint}{eprint}
}%
\BibSpec{book}{%
	+{}{\PrintAuthors}   {author}
	+{}{ \PrintDateB}    {date}
	+{}{ \itcaps}        {title}
	+{}{ \PrintEdition}  {edition}
	+{}{ \oparen}        {place}
	+{:}{ \cparen}       {publisher}
	+{}{ pp \eprintpages}{pages}
}%
\BibSpec{collection.article}{%
	+{}{\PrintAuthors}       {author}
	+{}{ \PrintDateB}        {date}
	+{}{ }                   {title}
	+{}{ \itcaps}            {booktitle}
	+{}{ ed \PrintEditorsIOP}{editor}
	+{}{ \oparen}            {place}
	+{:}{ \cparen}           {publisher}
	+{}{ pp \eprintpages}    {pages}
}%
%%% End bibliography format.
\bib{AW}{article}{
	author={Archer, F},
	author={Williams, R M},
	title={The Turaev-Viro state sum model and three-dimensional quantum gravity},
	journal={Physics Letters B},
	year={1991},
	volume={273},
	pages={438\ndash 44},
}
\bib{CY-BF}{article}{
	author={Baez, J C},
	title={4-dimensional $BF$ theory as a topological quantum field theory},
	journal={Letters in Mathematical Physics},
	year={1996},
	volume={38},
	pages={129\ndash 43},
	eprint={arXiv:q-alg/9507006},
}
\bib{Baez-spinfoam}{article}{
	author={Baez, J C},
	title={Spin foam models},
	journal={Classical and Quantum Gravity},
	year={1998},
	volume={15},
	pages={1827\ndash 58},
	eprint={arXiv:gr-qc/9709052},
}
\bib{Baez}{article}{
	author={Baez, J C},
	title={An introduction to spin foam models of quantum gravity and $BF$
		theory},
	journal={Lecture Notes in Physics},
	year={2000},
	volume={543},
	pages={25\ndash 94},
	eprint={arXiv:gr-qc/9904025},
}
\bib{BC-pos}{article}{
	author={Baez, J C},
	author={Christensen, J D},
	title={Positivity of spin foam amplitudes},
	journal={Classical and Quantum Gravity},
	year={2002},
	volume={19},
	pages={2291\ndash 306},
	eprint={arXiv:gr­qc/0110044},
}
\bib{BCE}{article}{
	author={Baez, J C},
	author={Christensen, J D},
	author={Egan, G},
	title={Asymptotics of $10j$ symbols},
	journal={Classical and Quantum Gravity},
	year={2002},
	volume={19},
	pages={6489\ndash 513},
	eprint={arXiv:gr-qc/0208010},
}
\bib{BCHT}{article}{
	author={Baez, J C},
	author={Christensen, J D},
	author={Halford, T R},
	author={Tsang, D C},
	title={Spin foam models of Riemannian quantum gravity},
	journal={Classical and Quantum Gravity},
	year={2002},
	volume={19},
	pages={4627\ndash 48},
	eprint={arXiv:gr-qc/0202017},
}
\bib{BC-riem}{article}{
	author={Barrett, J W},
	author={Crane, L},
	title={Relativistic spin networks and quantum gravity},
	journal={Journal of Mathematical Physics},
	year={1998},
	volume={39},
	pages={3296\ndash 302},
	eprint={arXiv:gr-qc/9709028},
}
\bib{BC-lor}{article}{
	author={Barrett, J W},
	author={Crane, L},
	title={A Lorentzian signature model for quantum general relativity},
	journal={Classical and Quantum Gravity},
	year={2000},
	volume={17},
	pages={3101\ndash 18},
	eprint={arXiv:gr-qc/9904025},
}
\bib{CFS}{book}{
	author={Carter, J S},
	author={Flath, D E},
	author={Saito, Masahico},
	title={The classical and quantum $6j$-symbols},
	year={1995},
	series={Mathematical Notes},
	volume={43},
	publisher={Princeton University Press},
	place={Princeton, New Jersey},
}
\bib{sgn-prob}{article}{
	author={Ceperley, D},
	author={Alder, B},
	title={Quantum Monte Carlo},
	journal={Science},
	year={1986},
	volume={231},
	pages={555\ndash 60},
}
\bib{CC}{article}{
	author={Cherrington, J W},
	author={Christensen, J D},
	title={Positivity in Lorentzian Barrett-Crane models of quantum
		gravity},
	journal={Classical and Quantum Gravity},
	year={2006},
	volume={23},
	pages={721\ndash 36},
	eprint={arXiv:gr-qc/0509080},
}
\bib{CE}{article}{
	author={Christensen, J D},
	author={Egan, G},
	title={An efficient algorithm for the Riemannian $10j$ symbols},
	journal={Classical and Quantum Gravity},
	year={2002},
	volume={19},
	pages={1184\ndash 93},
	eprint={arXiv:gr-qc/0110045},
}
\bib{CKY}{article}{
	author={Crane, L},
	author={Kauffman, L H},
	author={Yetter, D N},
	title={State-Sum Invariants of 4-Manifolds},
	journal={Journal of Knot Theory and Its Ramifications},
	year={1997},
	volume={6},
	pages={177\ndash 234},
	eprint={arXiv:hep-th/9409167},
}
\bib{CY}{article}{
	author={Crane, L},
	author={Yetter, D N},
	title={A categorical construction of 4D topological quantum field
		theories},
	booktitle={Quantum topology},
	editor={Kauffman, L H},
	editor={Baadhio, R A},
	year={1993},
	pages={120\ndash 30},
	publisher={World Scientific Press},
	place={Singapore},
}
\bib{DFKR}{article}{
	author={De Pietri, R},
	author={Freidel, L},
	author={Krasnov, K},
	author={Rovelli, C},
	title={Barrett-Crane model from a Boulatov-Ooguri field theory over a
		homogeneous space},
	journal={Nuclear Physics B},
	year={2000},
	volume={574},
	pages={785\ndash 806},
	eprint={arXiv:hep-th/9907154},
}
\bib{GFT}{article}{
	author={Freidel, L},
	title={Group field theory: an overview},
	journal={International Journal of Theoretical Physics},
	year={2005},
	volume={44},
	pages={1769\ndash 83},
	eprint={arXiv:hep-th/0505016},
}
\bib{KL}{book}{
	author={Kauffman, L H},
	author={Lins, S L},
	title={Temperley-Lieb recoupling theory and invariants of 3-manifolds},
	year={1994},
	series={Annals of Mathematics Studies},
	volume={134},
	publisher={Princeton University Press},
	place={Princeton, New Jersey},
}
\bib{KI}{article}{
	author={Kikuchi, M},
	author={Ito, N},
	title={Statistical dependence time and its application to dynamical
		critical exponent},
	journal={Journal of the Physical Society of Japan},
	year={1993},
	volume={62},
	pages={3052},
}
\bib{LB}{book}{
	author={Landau, D P},
	author={Binder, K},
	title={A guide to Monte Carlo simulations in statistical physics},
	year={2005},
	edition={2},
	publisher={Cambridge University Press},
	place={Cambridge},
}
\bib{qg-ref}{book}{
	author={Majid, S},
	title={Foundations of quantum group theory},
	year={2000},
	publisher={Cambridge University Press},
	place={Cambridge},
}
\bib{Metrop}{article}{
	author={Metropolis, N},
	author={Rosenbluth, A W},
	author={Rosenbluth, M N},
	author={Teller, A H},
	author={Teller, E},
	title={Equation of state calculations by fast computing machines},
	journal={Journal of Chemical Physics},
	year={1953},
	volume={21},
	pages={1087\ndash 92},
}
\bib{NR}{article}{
	author={Noui, K},
	author={Roche, P},
	title={Cosmological deformation of Lorentzian spin foam models},
	journal={Classical and Quantum Gravity},
	year={2003},
	volume={20},
	pages={3175\ndash 214},
	eprint={arXiv:gr-qc/0211109},
}
\bib{Perez}{article}{
	author={Perez, A},
	title={Spin foam models for quantum gravity},
	journal={Classical and Quantum Gravity},
	year={2003},
	volume={20},
	pages={R043\ndash 104},
	eprint={arXiv:gr-qc/0301113},
}
\bib{PeRo}{article}{
	author={Perez, A},
	author={Rovelli, C},
	title={A spin foam model without bubble divergences},
	journal={Nuclear Physics B},
	year={2001},
	volume={599},
	pages={255\ndash 82},
	eprint={arXiv:gr-qc/0006107},
}
\bib{Plebanski}{article}{
	author={Pleba\'nski, J F},
	title={On the separation of Einsteinian substructures},
	journal={Journal of Mathematical Physics},
	year={1977},
	volume={18},
	pages={2511\ndash 20},
}
\bib{PR}{article}{
	author={Ponzano, G},
	author={Regge, T},
	%This title from http://adsabs.harvard.edu/ is wrong.
	%title={On Racah coefficients as coupling coefficients for the vector
	%	space of Wigner operators},
	title={Semiclassical limit of Racah coefficients},
	editor={Bloch, F},
	editor={Cohen, S G},
	editor={De-Shalit, A},
	editor={Sambursky, S},
	editor={Talmi, I},
	booktitle={Spectroscopic and group theoretical methods in physics},
	pages={1\ndash 98},
	year={1968},
	publisher={North-Holland},
	place={Amsterdam},
}
\bib{BC-uniq}{article}{
	author={Reisenberger, M P},
	title={On relativistic spin network vertices},
	journal={Journal of Mathematical Physics},
	year={1999},
	volume={40},
	pages={2046\ndash 54},
	eprint={arXiv:gr-qc/9809067},
}
\bib{Smolin}{article}{
	author={Smolin, L},
	title={Linking topological quantum field theory and nonperturbative
		quantum gravity},
	journal={Journal of Mathematical Physics},
	year={1995},
	volume={36},
	pages={6417\ndash 55},
	eprint={arXiv:gr-qc/9505028},
}
\bib{Smolin-posCC}{eprint}{
	author={Smolin, L},
	title={Quantum gravity with a positive cosmological constant},
	date={2002},
	eprint={arXiv:hep-th/0209079},
}
\bib{TV}{article}{
	author={Turaev, V G},
	author={Viro, O Y},
	title={State sum invariants of $3$-manifolds and quantum
		$6j$-symbols},
	journal={Topology},
	year={1992},
	volume={31},
	pages={865\ndash 902},
}
\bib{Yetter}{article}{
	author={Yetter, D N},
	title={Generalized Barrett-Crane vertices and invariants of embedded
		graphs},
	journal={Journal of Knot Theory and Its Ramifications},
	year={1999},
	volume={8},
	pages={815\ndash 29},
	eprint={arXiv:math.QA/9801131},
}

\end{thebibliography}
\end{document}